\documentclass[aps,longbibliography,notitlepage,nofootinbib,11pt, floatfix,tightenlines,
]{revtex4-2}

\usepackage[utf8]{inputenc}
\usepackage[english,russian]{babel}
\usepackage{amsmath}
\usepackage{amssymb}
\usepackage{lipsum}
\usepackage{bm}
\usepackage{graphicx}
\usepackage{graphics}
\usepackage[dvipdf]{epsfig}
\usepackage[format=plain,labelfont={bf,small},textfont=small,justification=raggedright,singlelinecheck=false]{caption}
\usepackage{float}
\usepackage{wrapfig}
\usepackage{url}
\usepackage{epstopdf} 
\usepackage{xcolor}
\usepackage{tensor} 
\usepackage{setspace}

\newcommand{\bx}{{\mathbf x}}
\newcommand{\bX}{{\mathbf X}}

\newcommand{\bI}{{\mathbf I}}
\newcommand{\ba}{{\mathbf a}}
\newcommand{\bA}{{\mathbf A}}
\newcommand{\bn}{{\mathbf n}}
\newcommand{\bN}{{\mathbf N}}
\newcommand{\bQ}{{\mathbf Q}}
\newcommand{\bU}{{\mathbf U}}
\newcommand{\bV}{{\mathbf V}}

\begin{document}
\title{Dilation-invariant bending of elastic plates, and broken symmetry in shells}
\author{E. Vitral}
	\email{evitral@unr.edu}
\author{J. A. Hanna}
	\email{jhanna@unr.edu}
\affiliation{Department of Mechanical Engineering, University of Nevada,
    1664  N.  Virginia  St.\ (0312),  Reno,  NV  89557-0312,  U.S.A.}

\begin{abstract}
We propose bending energies for isotropic elastic plates and shells. 
For a plate, we define and employ a surface tensor that symmetrically couples stretch and curvature such that any elastic energy density constructed from its invariants is invariant under spatial dilations. 
This kinematic measure and its corresponding isotropic quadratic energy resolve outstanding issues in thin structure elasticity, including the natural extension of primitive bending strains for straight rods to plates, 
the assurance of a moment linear in the bending measure,
and the avoidance of induced mid-plane strains in response to pure moments as found in some commonly used analytical plate models. 
Our analysis also reveals that some other commonly used numerical models have the right invariance properties, although they lack full generality at quadratic order in stretch. 
We further extend our result to 
naturally-curved rods and shells, for which the pure stretching of a curved rest configuration breaks dilation invariance; the new shell bending measure we provide contrasts with previous \emph{ad hoc} postulated forms. 
The concept that unifies these theories is not dilation invariance, but rather through-thickness uniformity of strain as a definition of pure stretching deformations. Our results provide a clean basis for simple models of low-dimensional elastic systems, and should enable more accurate analytical probing of the structure of singularities in sheets and membranes. 
\end{abstract}

\date{\today}


\selectlanguage{english}\maketitle


\section{Introduction}

It is remarkable that after many decades of modern research on thin structures, there is not yet a consensus on a primitive covariant bending energy for simple isotropic elastic plates subject to combinations of stretching and bending deformations. 
Small-slope, F{\"{o}}ppl-von K{\'{a}}rm{\'{a}}n-like models are widely used for analytical approximations, but are appropriate only when changes in curvature are small.
Plates and shells form the basis for extensive modelling of low-dimensional systems such as elastic sheets and membranes in mathematics, physics, engineering, and biology \cite{Antman05, Villaggio97, Klein07, GuvenMuller08, Peterson09, Mellado11, Vernizzi11, Kim12science, GemmerVenkataramani13, NguyenSelinger17, Plucinsky18, KriegerDias19, Battista19-2, Stein-Montalvo19, Li19-2, Davidovitch19}. 
Exemplifying the lack of a universal standard, two commonly used bending energies in soft matter, namely the squared curvature and the ESK theory \cite{Efrati09jmps} based on Saint-Venant-Kirchhoff elasticity, display qualitatively opposite behaviors, either increasing or decreasing their curvature in response to simple loadings, and responding with extension or compression of a plate mid-plane to the application of pure moments \cite{IrschikGerstmayr09, OshriDiamant17, WoodHanna19}.
These effects will not exist in an appropriately primitive quadratic model in which the tangential force does not pick up extra bending contributions and the moment is linear in the corresponding strain measure. 
Given the interest in the soft matter community in the manner in which geometric rules enforce coupling between stretching and bending, particularly during the regularization of singular crumpling behaviors in sheets \cite{Witten07}, 
  it is imperative to avoid mixing these effects with the results of artificially introduced nonlinear couplings arising from constitutive choices. 
   
 Our intent in this brief note and a detailed companion paper \cite{vitralhannapt2} is to present new kinematic measures of plate and shell bending, and furthermore show that these results may be derived by dimensional reduction from an equally primitive bulk elastic energy.  
 The plate bending measure is a symmetric bilinear product of stretch and curvature whose immediate consequence is a dilation-invariant bending energy.
Both the direct construction and its more formal justification involve a recognition that strain measures linear in stretch are the appropriate primitive expansion variables for a small-strain elasticity theory, affording the simplicity of decoupling moments and plate mid-plane strains, as well as a linear relationship between moment and bending strain.  This linear response is the expected consequence of an energy quadratic in bending strain, but requires a careful choice of bending measure. 
While related ideas are known in the literature on the mechanics of rods \cite{IrschikGerstmayr09}, they have not been extended in any general way to two-dimensional objects and, we furthermore maintain, were not properly extended to curved one-dimensional objects until now.
A neo-Hookean thin body, as often used in simulations, will display the correct invariance properties but 
lacks one of two possible quadratic terms in its elastic energy. 
Curiously, our findings reveal that for naturally-flat plates, the commonly used discrete SN model \cite{SeungNelson88} also possesses the right invariance properties; despite purporting to represent a squared curvature energy, it actually corresponds to a one-parameter subset of our model, similar to a neo-Hookean body but with different proportions of invariants.

Our construction naturally extends and fully generalizes the primitive strains proposed over half a century ago by Antman and others \cite{Antman68-2,Reissner72,WhitmanDeSilva74} for straight rods.  Our plate bending measure agrees with those recently proposed in a restricted setting by Oshri and Diamant \cite{OshriDiamant17} and employed by Oshri \cite{Oshri19}. 
 An asymmetric referential form of this measure may be found in Atluri's \emph{tour de force} \cite{atluri1984alternate}, although this promising suggestion seems not to have been adopted elsewhere. 
 The special property of dilation invariance that we emphasize as a criterion for plate bending energies implies that dilations correspond purely to changes in the stretching energy of plates.  We demonstrate that this is a consequence of a preferred flat configuration; the simple example of a naturally-curved rod (beam) illustrates why and how shells violate this symmetry and require a different type of energy that is not dilation-invariant.  We propose an extension to shells that stands in direct contrast with dilation-invariant bending measures for one-dimensional curved rods and axisymmetric shells proposed by Antman and others \cite{Antman68-2,Reissner72,WhitmanDeSilva74, KnocheKierfeld11}, the two-dimensional shell bending measure proposed by Atluri \cite{atluri1984alternate} and also found in Pietraszkiewicz \cite{pietraszkiewicz2008determination}, and the na{\"{\i}}ve extension of discrete SN to shells.  This proposal is further justified in a companion paper \cite{vitralhannapt2} by a detailed derivation, which also recovers the classical unstretched (neutral, not mid-) surface of a curved body.
 
 The division of our work into two papers reflects the possibility of independent routes to the same conclusions. 
 The present note proposes two-dimensional energies and justifies them with physical arguments based on a definition of a pure stretching deformation of a thin body.
This direct construction does not rely on any choice of three-dimensional bulk energy.
The companion paper \cite{vitralhannapt2} employs a reduction from a particular quadratic model of bulk energy, and allows the same definition of pure stretching to arise naturally from the calculation. 
The arguments and assumptions of the two papers are distinct and, thus, the reader is free to find one or the other more palatable.
 We have attempted to present both approaches in a language intermediate between those of soft matter physicists and classical mechanicians.

 \section{Plate bending energy}

We use material coordinates $\eta^\alpha$, $\alpha \in \{1,2\}$, to parameterize rest (referential) and deformed (present) surfaces $\bX(\eta^\alpha)$ and $\bx(\eta^\alpha)$ in $\mathbb{E}^3$, with respective unit normals $\bN$ and $\bn$, tangents $\bA_\alpha = d_\alpha\bX$ and $\ba_\alpha = d_\alpha\bx$, and reciprocal tangents defined through the relations $\bA^\alpha\cdot\bA_\beta = \ba^\alpha\cdot\ba_\beta = \delta^\alpha_\beta$.
We consider the general form of a sum of stretching and bending energy densities defined per reference area (equivalent to per mass in the present context), 
\begin{align}\label{energy}
	\int\! \textrm{d}a\, J^{-1} \big[ \mathcal{W}_S\left(\bA_\alpha, \ba_\alpha\right) + \mathcal{W}_B\left( \bA_\alpha, \ba_\alpha, d_\beta\bA_\alpha, d_\beta\ba_\alpha \right) \big] \, , 
\end{align}
where $\textrm{d}a$ 
 is the present area form.  The referential area form $\textrm{d}A = \textrm{d}a\, J^{-1}$, where $J$ 
 can be computed as the ratio of present to referential metric determinants.  
The quantities listed in \eqref{energy} are closely related to the first and second derivatives of $\bx$ and $\bX$, and thus the symmetric tensors of the metric and curvature of both configurations. 
The present and referential metric components are respectively $a_{\alpha\beta} = \ba_\alpha\cdot\ba_\beta$ and $A_{\alpha\beta} = \bA_\alpha\cdot\bA_\beta$, while the inverses  are $a^{\alpha\beta} = \ba^\alpha\cdot\ba^\beta$ and $A^{\alpha\beta} = \bA^\alpha\cdot\bA^\beta$, taking care to note that indices should only be raised and lowered with the corresponding metric. 
One can think either in terms of first derivatives of the normals or second derivatives of the tangents, as the components of the curvature tensor $\mathbf{b} = b_{\alpha\beta}\ba^\alpha\ba^\beta = -\nabla\bn = -d_\alpha\bn \ba^\alpha$ are given by $b_{\alpha\beta} = d_\beta \ba_\alpha\cdot\bn = -\ba_\alpha \cdot d_\beta\bn$. 
For a plate, 
the curvature tensor of the reference configuration  $\mathbf{\bar b} = d_\beta \bA_\alpha\cdot\bN \bA^\alpha\bA^\beta$ vanishes.

The present note is concerned solely with the bending content $\mathcal{W}_B$ of \eqref{energy}. 
To define our bending measure, we require the point-wise decomposition of the surface deformation gradient into a (3D) rotation tensor $\bQ \in$ SO(3) and the (2D) symmetric right (referential) $\bU$ or left (present) $\bV$ surface stretch tensor 
\begin{align}
	\ba_\alpha\bA^\alpha = \bQ\cdot\bU = \bV\cdot\bQ \, .
\end{align}
This is a justifiable \cite{pietraszkiewicz2008determination} abuse of notation, as while $\bQ$ is fully three-dimensional, the tensors $\bU$ and $\bV$ are restricted to the reference and present surfaces, respectively, and $\bQ\cdot\bN = \bN\cdot\bQ^\top = \bn$ and $\bQ^\top\cdot\bn=\bn\cdot\bQ =\bN$. 
The stretches are tensors encoding the ratio of present length to rest length; note that when there is no surface strain, so colloquially ``no stretching'', the stretch tensors are $\bU = \bA_\alpha\bA^\alpha$ and $\bV = \ba_\alpha\ba^\alpha$ with principal stretches of unity, not zero.
The natural strain measure here is the surface Bell strain $\bV - \ba_\alpha\ba^\alpha$.
 The transpose of $\mathbf{Q}$ is its inverse, and thus we have $\mathbf{Q}^\top\cdot\mathbf{Q} = \bA_\alpha\bA^\alpha + \bN\bN = \bI =  \ba_\alpha\ba^\alpha + \bn\bn = \mathbf{Q}\cdot\mathbf{Q}^\top$, from whence $\bU\cdot\bU = a_{\alpha\beta}\bA^\alpha\bA^\beta$ and $\bV\cdot\bV = A^{\alpha\beta}\ba_\alpha\ba_\beta$.
While the stretch and curvature tensors are all symmetric, dot products between them are not, reflecting the independence of principal directions of stretching and bending.

\begin{figure}[h]
	\includegraphics[width=3in]{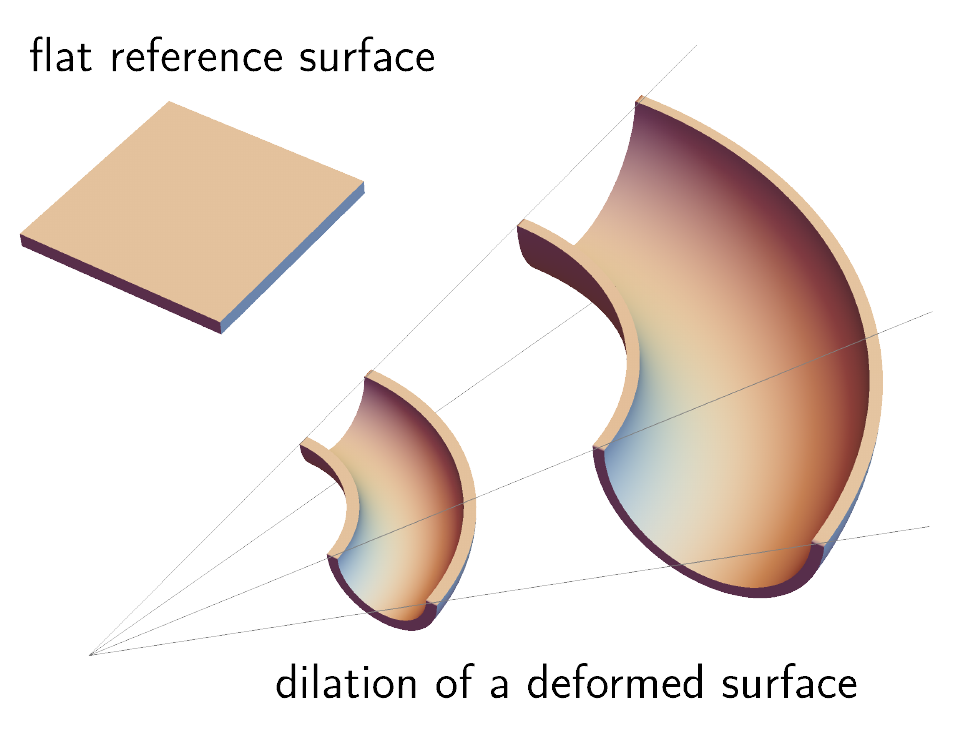}
	\captionsetup{margin=0.5in}
	\caption{Schematic of a dilation of a deformed surface whose reference configuration is flat.}\label{dilation}
\end{figure}

Our central result for plates is the tensor measure of bending
\selectlanguage{russian}
\begin{align}
	\text{\Large{\bf{\cyrl}}} = \mathrm{sym} \left( \bV \cdot \mathbf{b} \right) = \tfrac{1}{2}\left( \bV \cdot \mathbf{b} + \mathbf{b} \cdot \bV \right) \, , \label{lurie}
\end{align}
\selectlanguage{english}
which reduces to the curvature tensor $\mathbf{b}$ when the in-surface strains vanish. 
Equivalently one can couple the referential quantities $\bU = \bQ^\top\cdot\bV\cdot\bQ$ and $\bQ^\top\cdot\mathbf{b}\cdot\bQ$, or use the fact that $d_\alpha\bn\bA^\alpha = \bar\nabla\bn = -\mathbf{b}\cdot\bV\cdot\bQ$ to interpret \eqref{lurie} in terms of $\bQ^\top$ and the referential gradient of the normal $\bar\nabla\bn$.  An asymmetric referential form of \eqref{lurie} was proposed
by Atluri \cite{atluri1984alternate}; both agree with established one-dimensional primitive measures \cite{Antman68-2,Reissner72,WhitmanDeSilva74, IrschikGerstmayr09, OshriDiamant17}. 
The surface tensor \eqref{lurie} is invariant under spatial dilations, which take any surface $\bx \rightarrow D\left(\bx - \bx_c\right)$. Thus, $\ba_\alpha \rightarrow D\ba_\alpha$ and therefore $\ba^\alpha \rightarrow D^{-1}\ba^\alpha$, while the normal and rotation are conserved, $\bn \rightarrow \bn$ and $\bQ\rightarrow\bQ$, so that $\bV \rightarrow D\bV$ and $\mathbf{b}\rightarrow  D^{-1}\mathbf{b}$ and their product is conserved.
Figure \ref{dilation} illustrates the situation under discussion, in which a dilation may be superposed on any previous stretching and bending deformations of a flat surface. 

Defining two convenient independent invariants of this symmetric tensor
\selectlanguage{russian}
$i_1 = \mathrm{Tr} \, \text{\Large{\bf{\cyrl}}}$ and 
$i_2 = \tfrac{1}{2}\left[ \left(\mathrm{Tr} \, \text{\Large{\bf{\cyrl}}}\right)^2 - \mathrm{Tr}\left(\text{\Large{\bf{\cyrl}}}\cdot\text{\Large{\bf{\cyrl}}}\right) \right] = \mathrm{Det} \, \text{\Large{\bf{\cyrl}}}$
\selectlanguage{english}
which, when the strain vanishes, reduce respectively to twice the mean, and the Gau{\ss}ian, curvature, the most general isotropic quadratic bending energy is
\begin{align}\label{bendingenergy}
	\mathcal{W}_B = 
	\tilde c_1 i_1^2 + 
	\tilde c_2 i_2 \, .
\end{align}
In the present note we seek to justify the use of \eqref{lurie} and thus \eqref{bendingenergy} through physical arguments.  A companion paper \cite{vitralhannapt2} derives stretching and this bending energy through reduction of a three-dimensional isotropic quadratic-stretch energy \cite{VitralHanna21}. 
We also find that using a three-dimensional neo-Hookean energy results in only one quadratic invariant, namely \selectlanguage{russian}$\mathrm{Tr}\left(\text{\Large{\bf{\cyrl}}}\cdot\text{\Large{\bf{\cyrl}}}\right) = i_1^2 - 2i_2$.\selectlanguage{english}  
Similarly, the continuum limit of the discrete SN model provides a single bending invariant;
 as normals are unaffected by dilation, any such scheme that uses only operations on adjacent normals to define a bending energy will have the appropriate invariance properties for a plate. 
This is somewhat ironic, as Seung and Nelson incorrectly assumed that they were discretizing a per-area geometric curvature energy \cite{Willmore65, Helfrich73} rather than a per-mass energy bilinear in stretch and curvature.
For a clear exposition of several possible discrete curvature definitions, we recommend \cite{Vouga14}.  Recall that, in contrast to a fluid membrane or other geometric energy, any sensible elastic energy should be defined per-mass rather than per-area \cite{GreenZerna92, Eringen67, Peterson09, Hanna19}. 
This surprising but fortunate result implies that both simulations of the discrete two-dimensional SN model, and very likely any three-dimensional finite element simulations of a neo-Hookean model for thin plates, will display the desirable properties of our proposed bending measure.  However, both the stretching and bending responses of these one-parameter models will lack some important features of a more general two-parameter quadratic model; it is impossible to choose the Poisson's ratio in the linear limit, and the models fail to capture many qualitative nonlinear behaviors of soft materials under simple loadings \cite{HorganSmayda12invariant}, 
while the typical extensions to Mooney-Rivlin or similar models will not preserve the properties we seek.
Furthermore, as discussed below, a na{\"{\i}}ve extension of an SN model to shells will incorrectly continue to preserve dilation invariance, and so not behave like either our shell energy or a neo-Hookean thin body.

It is interesting to note that the referential-weighted determinant of \eqref{lurie} is a purely geometric quantity, the integral of Gau{\ss}ian curvature $\int\! \textrm{d}A\, i_2 = \int \textrm{d}A\, \mathrm{Det}\, \bV\, \mathrm{Det}\, \mathbf{b} = \int\! \textrm{d}a\, \mathrm{Det}\, \mathbf{b}$, 
which is equal to a boundary term plus a topological invariant, and therefore constant for any closed surface.
This is curious because, again, we don't expect an elastic energy to be geometric.
It further implies that SN or neo-Hookean models may sufficiently capture the behavior of closed surfaces composed of quadratic-stretch plate-like material, however it is more likely that a study of a closed surface will be concerned with a shell energy instead of a plate energy.

We may contrast the bending energy \eqref{bendingenergy} with commonly used energies similarly constructed from invariants of other bending measures.
The measure \selectlanguage{russian}$\text{\Large{\bf{\cyrl}}}$\selectlanguage{english} and by extension $i_1$, $i_2$, and their integrals using the referential measure $\textrm{d}A$ are all dilation-invariant.  This means that a dilation is a pure stretching deformation that affects only the stretching content $\mathcal{W}_S$. 
The squared curvature energy uses the curvature tensor $\mathbf{b}$ and decreases under dilations, as these decrease geometric curvatures. 
The ESK energy uses the unnamed tensor $b_{\alpha\beta}\bA^\alpha\bA^\beta$ and increases under dilations \cite{WoodHanna19}. 
Note also that the per-area geometric energy $\int\! \textrm{d}a \left(\mathrm{Tr}\, \mathbf{b}\right)^2$, suitable for fluid films, is conformally invariant up to a boundary term \cite{White73}, 
while its proper per-mass elastic counterpart $\int\! \textrm{d}A \left(\mathrm{Tr}\, \mathbf{b}\right)^2$ forms part of the squared curvature energy and has no notable invariance properties; the other part 
$\int\! \textrm{d}A\, \mathrm{Det}\, \mathbf{b}$ is not a boundary/topological term like its geometric counterpart.

In the limit of a mid-surface isometry, all of these energies degenerate to the same value, although not the same slope with respect to strain \cite{WoodHanna19}.  However, this is a highly restricted situation.  Isometries are not generically possible under many loading or confinement conditions, and most importantly even when an isometry is possible it will not be adopted by the plate even under very simple boundary conditions unless the force and moment have appropriately simple relationships with the stretching and bending measures \cite{OshriDiamant17, WoodHanna19}. The oft-assumed paradigm of first minimizing stretching energy by finding an isometry, then subsequently minimizing bending energy among a class of isometries, might be appropriate for free boundaries but has been demonstrated not to work in general for squared curvature or ESK energies \cite{OshriDiamant17, WoodHanna19}.

We reserve the derivation and presentation of full field equations and boundary conditions associated with the energy \eqref{energy} for a companion paper \cite{vitralhannapt2}, but here demonstrate linearity of the bending response. 
That is, a quadratic strain energy will give rise to a moment linear in the strain measure.  Many seemingly simple bending measures do not have this property. 
This is a subtle point discussed in detail in the engineering mechanics literature by Irschik and Gerstmayr \cite{IrschikGerstmayr09} and from a physics perspective by Oshri and Diamant \cite{OshriDiamant17}, both in a one-dimensional setting. 
The variation\footnote{Variations of present quantities are simplest to derive in terms of an integral over present area, but can be easily translated into referential form via Piola identities such as $\nabla_\alpha\left[J^{-1}()^\alpha\right] = J^{-1}\bar\nabla_\alpha()^\alpha$ or $\bar\nabla_\alpha\left[J()^\alpha\right] = J\nabla_\alpha()^\alpha$.}   of an isotropic quadratic energy density akin to \eqref{bendingenergy} from invariants of a symmetric tensor ${\mathbf{S}}$ will take the form 
$\left[ \left(2\tilde c_1+\tilde c_2\right) S^\gamma_\gamma \delta^\alpha_\beta  - \tilde c_2S^\alpha_\beta \right] \delta S_\alpha^\beta$, bilinear in the tensor and its variation.
 The variation $\delta S_\alpha^\beta$ is then related to variation of position and its derivatives, and will contribute both to the stress and the boundary stress and moment, the last coming only from terms involving second derivatives of position. The moment is the part of the boundary term conjugate to the variation of the angles of the tangent plane \cite{wisniewski1998shell, IrschikGerstmayr09, OshriDiamant17}.  Thus, we want this part of the variation of the measure to have no dependence on the measure. 
For our measure, it suffices to consider one asymmetric half of the variation, $\delta\left(V^{\beta\gamma}b_{\gamma\alpha}\right)$.  Using the fact \cite{Hanna19} that $\delta b_{\gamma\alpha} = \nabla_\alpha\nabla_\gamma\delta\bx\cdot\bn$ and rewriting $\nabla_\gamma\delta\bx = \delta\ba_\gamma = \delta\left(V_\gamma^\delta\tilde\ba_\delta\right)$, where
$\tilde\ba_\delta$ is an unstretched but rotated referential tangent whose variation represents a change in angle, we extract the quantity conjugate to $\delta\tilde\ba_\delta$.  This moment is linear in the measure \selectlanguage{russian}\text{\Large{\bf{\cyrl}}},\selectlanguage{english} as a term such as $V^{\beta\gamma}V_\gamma^\delta = A^{\beta\delta}$ is independent of stretch or curvature, indeed entirely independent of the deformation. 
By contrast, the variation of the squared curvature energy $\delta\left(a^{\beta\gamma}b_{\gamma\alpha}\right)$ provides an inverse stretch term $V^{\delta\beta}$ to multiply the term linear in the measure $\mathbf{b}$, while that of the ESK energy $\delta\left(A^{\beta\gamma}b_{\gamma\alpha}\right)$ provides a (physical) stretch term $A^{\beta\gamma}V^\delta_\gamma$ to multiply the term linear in the measure $b_{\alpha\beta}\bA^\alpha\bA^\beta$.

\begin{figure}[h]
	\includegraphics[width=6in]{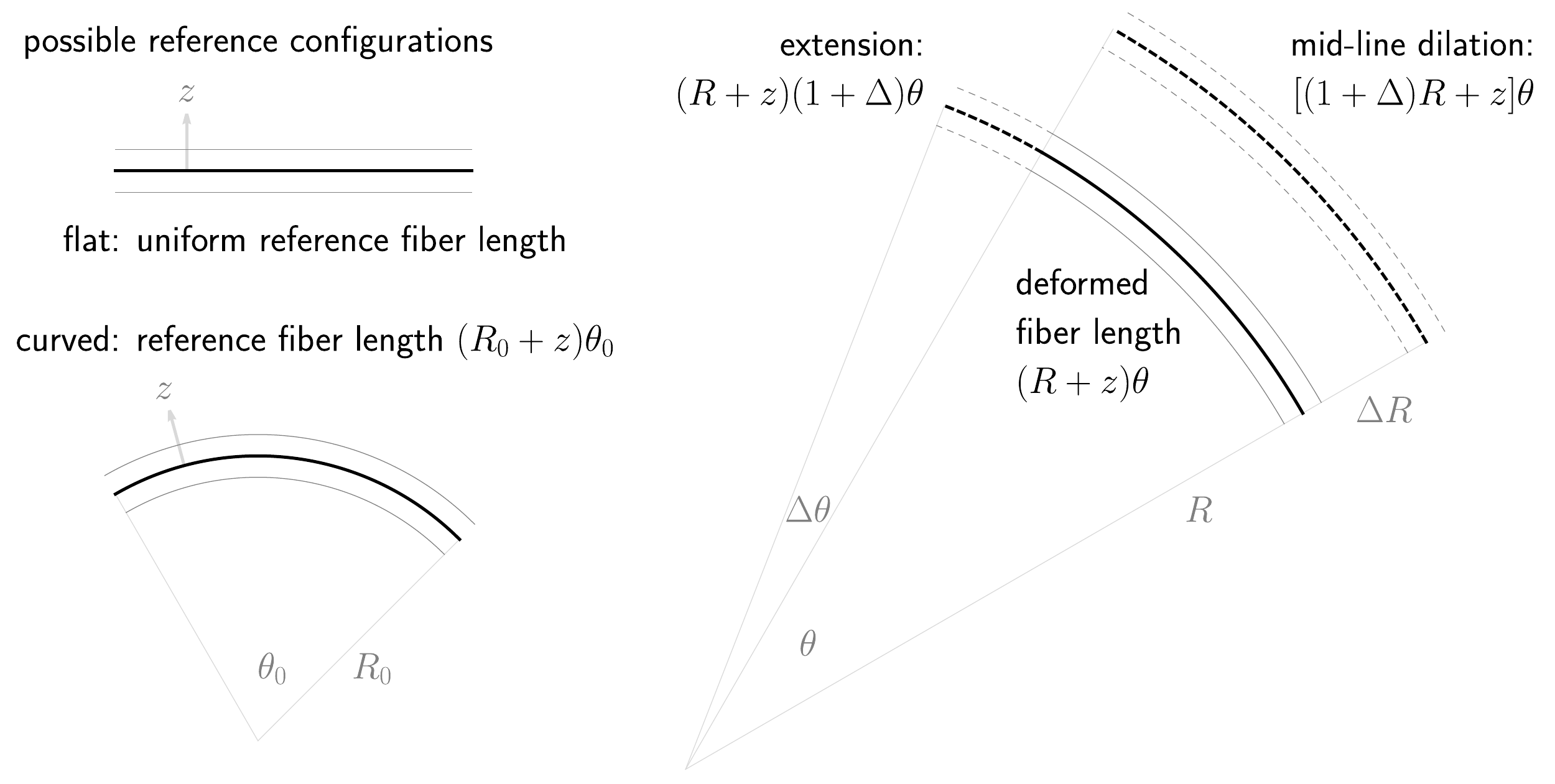}
	\captionsetup{margin=0.5in}
	\caption{Transformations of a deformed one-dimensional plate or shell.  Pure stretches involve uniform stretch across the thickness direction $z$.  Extensions are only pure stretches if performed at the referential curvature.  Dilations are only pure stretches if the reference configuration is flat.}\label{transformations}
\end{figure}

\section{Pure stretching, and shell bending energy}

For a physical understanding of why dilations correspond to pure stretching of a plate, it is best to consider both naturally-flat plates and naturally-curved shells together.  We illustrate the idea in Figure \ref{transformations} using a one-dimensional plate or shell (rod) with radius $R_0$ in its rest configuration. 
The central unifying idea is not actually dilation invariance, but the idea of tangential extension and associated through-thickness uniformity of stretch. 
The natural definition of a pure stretching deformation of a thin body is the absence of a gradient in tangential stretch (and by extension, higher-order strains such as the metric) across the thickness of the body; by contrast, a simple bending deformation would provide a through-thickness linear variation in stretch (quadratic in the metric).
Consider the stretch of the rod mid-line and of other material fibers offset on either side of the mid-line by a distance $z$; although these latter fibers will contract towards the mid-line by a factor related to Poisson's ratio, for simplicity we may consider distortions of a thin body maintaining constant thickness. 
Consider a deformed piece of surface of radius $R$ subtending an angle $\theta$.  Extending this shape along itself, maintaining its radius of curvature, extends tangential fibers of length $(R+z)\theta$ to length $(R+z)(1 + \Delta)\theta$, which additional change in length $(R+z)\Delta\theta$ should be compared with the reference length $(R_0+z)\theta_0$.  If the curvature is the reference curvature, $R=R_0$ and the additional stretch is $(1+\Delta)\theta/\theta_0$ independent of position $z$.
Thus, extension of a flat or curved rod along its referential shape, a tangential deformation in the reference configuration, is a pure stretching deformation. 
In the special case of a flat (straight) rod, this is just uniform extension of straight lines on, or offset from, the referential flat mid-line.  
This uniform extension of curves across the thickness is achieved by dilations of any deformed surface,\footnote{Note that it is the surface alone, and not the thin body surrounding it, that is undergoing dilation.} which extend fibers of length $(R+z)\theta$ to length $\left[(1+\Delta)R+z\right]\theta$.  The uniform extension by $\Delta R \theta$ corresponds to a uniform through-thickness stretch only if the reference fiber lengths are also uniform in $z$, as with a flat rod.

As tangential extension along a curved referential shape is a pure stretch for a shell, pure stretching does not correspond to dilations and the appropriate bending energy is not dilation-invariant.
Considering our definition of pure stretching, or the detailed derivation in the companion paper \cite{vitralhannapt2}, provides us with the shell bending measure 
$\mathrm{sym}\left[ \mathbf{b}\cdot\bV - \bQ\cdot\mathbf{\bar b}\cdot\bU\cdot\bQ^\top \right]$ =
$\mathrm{sym}\left[ \left(\mathbf{b} - \bQ\cdot\mathbf{\bar b}\cdot\bQ^\top\right)\cdot\bV \right]$, 
which clearly inherits the linearity property of the plate measure, as the only portion contributing to the moment is the variation of $\mathbf{b}$.
Equivalently one can use the referential tensor $\mathrm{sym}\left[ \bQ^\top\cdot\mathbf{b}\cdot\bV\cdot\bQ - \mathbf{\bar b}\cdot\bU \right]$ = $\mathrm{sym}\left[ \left(\bQ^\top\cdot\mathbf{b}\cdot\bQ - \mathbf{\bar b}\right)\cdot\bU \right]$.  While these forms appear cumbersome at first glance, they can be inferred by the requirement to subtract tensors of like type. 
For the simple example of a one-dimensional body, we obtain a bending energy quadratic in a single quantity
\begin{align}
	\lambda\left(\kappa - \bar\kappa \right) \, , \label{1dbending}
\end{align}
where in terms of a single material coordinate $l$ along the curve, $\lambda = V_l^l = a_{ll}V^{ll}$ is the stretch while $\kappa = b_l^l = a^{ll}d_l^2\bx\cdot\bn$ and $\bar\kappa = A^{ll}d_l^2\bX\cdot\bN$ are the curvatures of the deformed and rest configurations of the rod. 
Clearly the form of \eqref{1dbending} is such that a tangential extension of a curve, maintaining the reference curvature, keeps the bending energy zero.  This is in contrast to the \emph{ad hoc} postulated dilation-invariant form $\lambda\kappa - \bar\kappa$ found throughout the literature \cite{Antman68-2,Reissner72,WhitmanDeSilva74, atluri1984alternate, pietraszkiewicz2008determination, KnocheKierfeld11} or a na{\"{\i}}ve extension of the discrete SN model.
Both forms will provide a linear moment, but differences will show up in the contribution to the force, which affects whether a pure moment will couple to stretching deformations, as discussed in the companion paper \cite{vitralhannapt2}. 
Note that tangential deformations of material on a surface are only ``pure stretching'' if performed in a state of reference curvature.  In an arbitrarily deformed configuration, tangential deformations change the bending content through the stretch tensor. 

\section{Discussion and conclusions}

Numerical implementation may be accomplished in one of two ways. Stretch or equivalently Biot/Bell strain is the natural language for linear elements in the bead-spring molecular dynamics simulations often preferred by physicists studying soft matter mesostructures. 
Techniques from computer graphics 
\cite{chen2018physical} 
may be employed to define discrete fundamental forms or, more directly, the stretch tensor and referential gradients of the normal on triangular meshes. 
Alternately, finite element methods exist that explicitly deal with 
 the rotation and stretch fields \cite{atluri1984alternate, wisniewski1998shell}, while 
 recently developed expressions \cite{VitralHanna21, vitralhannapt2} can provide 
 all of the relevant fields in terms of metrics and curvatures, and therefore derivatives of position. 

The primitive bending strains we have introduced provide a new basis for analytical modeling of plates and shells as well as curved rods, bringing these under the same framework as was established for straight rods half a century ago.
We are also reassured that studies probing singular structures in naturally-flat elastic sheets by employing a discrete SN model \cite{Witten07}, while possibly missing some elastic effects found in more general quadratic models, have likely not introduced artificial constitutive stretch-bend coupling and are thus on a similar footing as simulations of neo-Hookean materials.

\section*{Acknowledgments}

This work was supported by U.S. National Science Foundation grant CMMI-2001262.  
We thank E. G. Virga for extensive detailed discussions and for sharing notes on related work.  
We also acknowledge helpful discussions with S. Cheng, P. Plucinsky, and E. Vouga.

\bibliographystyle{unsrt}

\end{document}